\documentclass[a4paper]{jpconf} 
\usepackage{graphicx}
\begin{document}
\title{The search of higher multipole radiation in gravitational waves from compact binary coalescences by a minimally-modelled pipeline}

\author{O. Halim$^{1}$, G. Vedovato$^2$, E. Milotti$^{3,1}$, G. A. Prodi$^{4,5}$, S. Bini$^{6,5}$, M. Drago$^{7,8}$, V. Gayathri$^9$, C. Lazzaro$^{10,2}$, D. Lopez$^{11}$, A. Miani$^{6,5}$, B. O’Brien$^9$, F. Salemi$^{6,5}$, M. Szczepanczyk$^9$, S. Tiwari$^{11}$, A. Virtuoso$^{3,1}$, S. Klimenko$^9$}

\address{$^1$INFN Sezione di Trieste, I-34127 Trieste, Italy\\
$^2$INFN, Sezione di Padova, I-35131 Padova, Italy\\
$^3$Universit\`{a} di Trieste, Dipartimento di Fisica, I-34127 Trieste, Italy\\
$^4$Universit\`{a} di Trento, Dipartimento di Matematica, I-38123 Povo, Trento, Italy\\
$^5$INFN, TIFPA, I-38123 Povo, Trento, Italy\\
$^6$Universit\`{a} di Trento, Dipartimento di Fisica, I-38123 Povo, Trento, Italy\\
$^7$Universit\`{a} di Roma La Sapienza, I-00185 Roma, Italy\\
$^8$INFN, Sezione di Roma, I-00185 Roma, Italy\\
$^9$University of Florida, Gainesville, FL 32611, USA\\
$^{10}$Universit\`{a} di Padova, Dipartimento di Fisica e Astronomia, I-35131 Padova, Italy\\
$^{11}$Physik-Institut, University of Zurich, Winterthurerstrasse 190, 8057 Zurich, Switzerland
}

\ead{odysse.halim@ts.infn.it}

\begin{abstract}
The coherent WaveBurst (cWB) pipeline implements a minimally-modelled search to find a coherent response in the network of gravitational wave detectors of the LIGO-Virgo Collaboration in the time-frequency domain. In this manuscript, we provide a timely introduction to an extension of the cWB analysis to detect spectral features beyond the main quadrupolar emission of gravitational waves during the inspiral phase of compact binary coalescences; more detailed discussion will be provided in a forthcoming paper \cite{hom}. The search is performed by defining specific regions in the time-frequency map to extract the energy of harmonics of main quadrupole mode in the inspiral phase. This method has already been used in the GW190814 discovery paper (Astrophys. J. Lett. \textbf{896} L44). Here we show the procedure to detect the (3, 3) multipole in GW190814 within the cWB framework. 
\end{abstract}


\section{Introduction}
Asymmetric binary blackhole systems are predicted to emit gravitational waves (GWs) with higher modes (HMs) in addition to the quadrupole \cite{LIGOScientific:2020stg}. The analyses of recent compact binary coalescence (CBC) signals by the LIGO-Virgo Collaboration demonstrated the existence of HMs \cite{LIGOScientific:2020stg,Abbott:2020khf}. The more sophisticated waveform models are required to describe binary systems including HMs, which are used for matched filter analysis pipelines. However, minimally-modelled burst algorithms, such as cWB can also detect this effect \cite{Klimenko_2016,Drago:2020kic}.

The cWB works at the level of time-frequency analysis. In terms of HM search from CBCs, it looks for coherent excess power in chirp-like regions corresponding to different HMs. This HM search strategy involving cWB, in some sense, is similar to an alternative method in \cite{Roy:2019phx}, which can be used to compare cWB reconstructions with the estimates from Bayesian inference method \cite{PhysRevD.100.042003,Payne:2019wmy}.

This proceeding is organised as follows. Sec.~\ref{sec:waveform} will explain the procedure. Sec.~\ref{sec:performances} shows the implementation on GW190814. In the end, we conclude everything in Sec.~\ref{sec:remarks}.


\section{The procedure: waveform residual energy}
\label{sec:waveform}

In GW analysis with cWB pipeline \cite{Klimenko_2016,Drago:2020kic}, waveform reconstructions are done thanks to the the discrete Wilson--Daubechies--Meyer (WDM) transform \cite{Necula:2012zz}. The pipeline first decompose the GW signals with this WDM in order to produce time-frequency maps of the signals. The time-frequency pixels are selected by retaining only a fixed fraction of them choosing those above a specified excess network energy. Moreover, cWB estimates the coherence among GW detectors by the maximization of the constrained likelihood \cite{Klimenko_2016}. These coherent wavelet pixels provide a GW waveform reconstruction, as a point estimate in the time domain.

In this proceeding, we summarise our recent paper \cite{hom} that has been submitted to CQG, in which we focus on the procedure to specifically detect the HMs. We use two compatible waveform models, in which the only difference is whether there is HM content or not. All the waveforms are whitened by cWB pipeline \cite{PhysRevD.100.042003}. We define ``on-source'' and ``off-source'' data, where on-source means the data contain the GW signal and off-source do not. These off-source data provide independent noise instances. In this case, we can assess the effect of noise with no assumption of the noise statistics except stationarity.

Our idea is to compare the cWB reconstructions with the model waveforms (with or without HMs) from the Bayesian inference methods \cite{Veitch_2015,Ashton_2019}. Thus, we define {\it waveform residual energy}, $E_\mathrm{res}$:
\begin{equation}
\label{Eres}
 E_\mathrm{res} =  \sum_{k=1}^\mathrm{det} \; \sum_{i\in\langle \mathrm{pixels} \rangle} (w_{k}^\mathrm{cWB}[i] - w_{k}^\mathrm{model}[i])^2,
\end{equation}
where  $w_{k}^\mathrm{cWB}[i]$ and $w_{k}^\mathrm{model}[i]$ are the WDM transforms of the cWB reconstruction 
and of a waveform model. Here, $k$ is the detector index and $i$ is the WDM pixel index in the time-frequency map (more details are in c.f Sec.~III.A of \cite{PhysRevD.100.042003}).

We test the consistency of cWB reconstruction comparing with the Bayesian estimation by calculating the residual energy $E_\mathrm{res}^\mathrm{(on-source)}$ in the on source reconstruction versus the maximum likelihood (MaxL) sample waveform from the Bayesian inference without HMs\footnote{From the Bayesian parameter estimate, we use the MaxL sample as the best estimate instead of \textit{maximum a posteriori} sample (MAP), which would be better motivated in a Bayesian perspective, because of the flat priors on specific regions in the parameter space. The MAP sample provides no real advantage when it is within the interior of these regions, and in fact it may produce worse results when it rails against their boundaries.}. The significance is evaluated by the empirical distribution of off-source injections from random samples of the posterior distribution into a wide set of off-source, equally spaced intervals.

The injected signals in off-source data are analysed and reconstructed by cWB and again, compared with their whitened version without HMs by $E_\mathrm{res}^\mathrm{(off-source)}$. Thus, the residual energies give us an empirical distribution either $E_\mathrm{res}^\mathrm{(on-source)}$ or $E_\mathrm{res}^\mathrm{(off-source)}$. From here, the \textit{p}-value can be calculated to test the hypothesis whether the injected waveform is in agreement with its cWB reconstruction.

The instantaneous frequency of the generic $(\ell, m)$ multipole\footnote{We use  $(\ell, m)$ as a shorthand for both $(\ell, m)$ and $(\ell, -m)$.} emitted by spinning, non-precessing black hole binaries, can be approximated by a scaling from the dominant $(2, 2)$ multipole \cite{Blanchet:2013haa,London:2017bcn}
\begin{equation}
\label{fmodes}
f_{\ell,m}(t) \approx \frac{m}{2} f_{22}(t).
\end{equation}
Therefore, we look for the presence of significant residual energy along ``slices'' of the time-frequency map; the slice is defined as  $f(t) = (\alpha\pm \delta \alpha) f_{22}(t)$, where  $\alpha$ is a non-negative real parameter~\cite{LIGOScientific:2020stg,Roy:2019phx} and $\delta\alpha$ determines the strip width, between a minimum and a maximum time. More details of the resolution $\delta\alpha$ can be seen in c.f. Fig.~4 of \cite{hom}.

\begin{figure*}[ht!]
\centering
\begin{centering}
\includegraphics[width=0.4\textwidth]{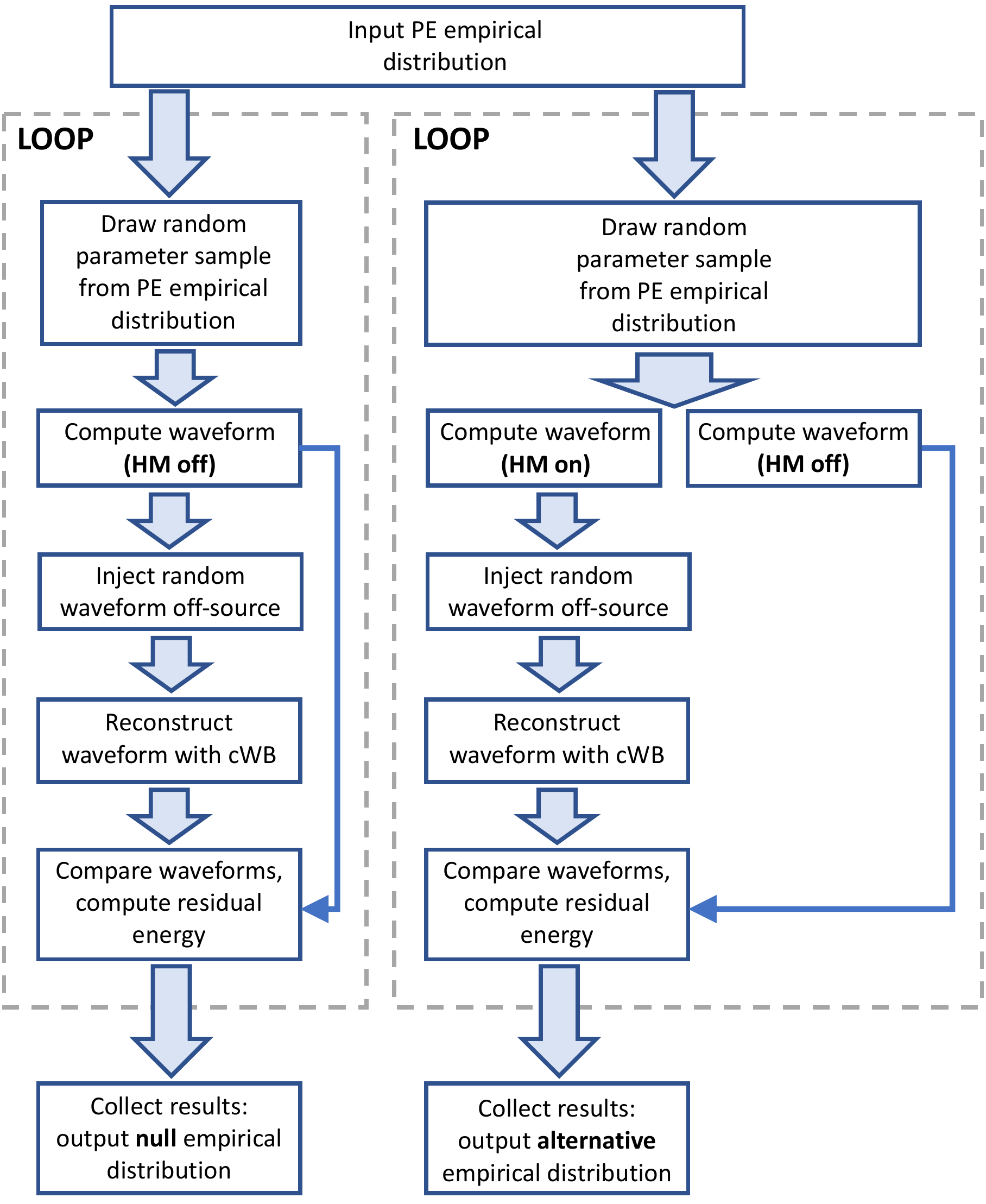}
\end{centering}	
\caption{\label{f:FlowChart} {Scheme of our strategy using Monte Carlo program to produce the null and alternative empirical distributions. Notice that the two different cWB reconstructions are both compared with the waveform \textbf{without} HMs (solid blue arrows). The Monte Carlo loop is repeated for $\sim$2000 off-source injections.} }
\end{figure*}

We calculate the significance of the residual energy by Monte Carlo simulations: random waveform samples from the Bayesian inference are injected into the off-source data both with and without HMs. For each case, we compute the residual energy. Thus, two empirical distributions are constructed: without HMs as the null hypothesis and with HMs as the alternative hypothesis. Our strategy can be seen step-by-step in the scheme of Fig \ref{f:FlowChart}.




\section{GW190814}
\label{sec:performances}

\begin{figure*}[ht!]
\centering
\includegraphics[clip,height=0.2\textheight]{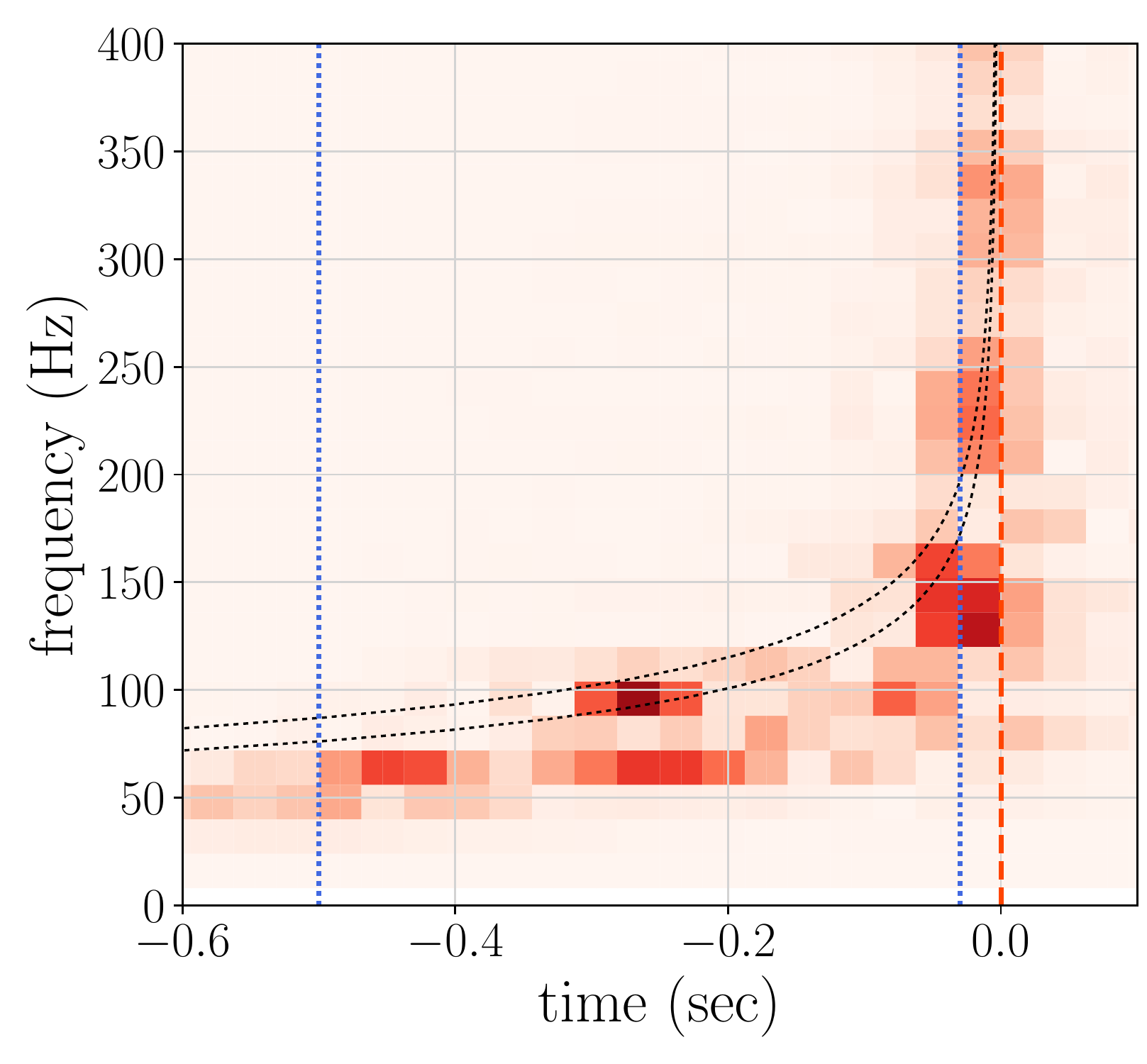} \hspace{0.02\textwidth}
\includegraphics[clip,height=0.2\textheight]{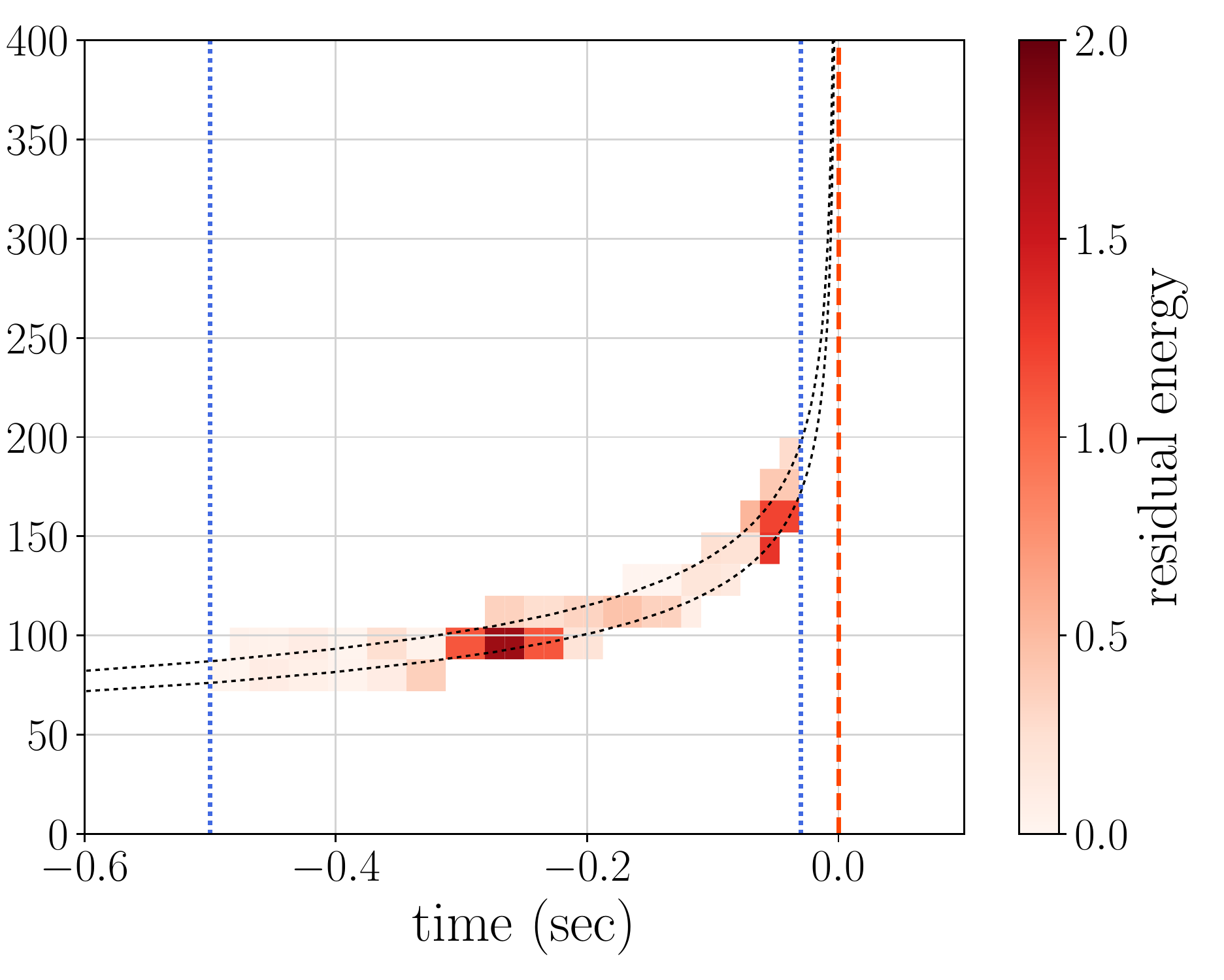} \\
\caption{\label{f:TFslice} \textbf{Left}: time-frequency map of $E_\mathrm{res}$ for the GW190814 with respect to the MaxL {\tt SEOBNRv4\_ROM} waveform, mapped into the LIGO Livingston detector by the cWB with resolution $dt = 1/32$~s and $df = 16$~Hz. \textbf{Right}: the time-frequency map of the time-frequency slice with $\alpha=1.5$ (meaning the  $(3,3)$ HM). These pixels are used to evaluate the total $E_\mathrm{res}(\alpha; \delta \alpha, \Delta t, \delta t, df)$ in the time-frequency slice. The red vertical line shows the merger time\footnote{GPS time: 1249852257.0154 s.} from the MaxL {\tt SEOBNRv4\_ROM} waveform. The dotted blue vertical lines are the considered time-frequency slice,  $[t_\mathrm{merger} - 0.5$~s$, t_\mathrm{merger} - 0.03$~s$]$. The black dotted curves are the limits of the time-frequency slice $[\alpha- 0.1, \alpha+ 0.1] \times f_{22}(t)$ with $\alpha = 1.5$.}

\end{figure*}

We highlight several results related to GW190814. For more details, the reader is advised to see \cite{hom}. First of all, let us discuss the left panel of Fig.~\ref{f:TFslice} for GW190814 analysis by cWB. This plot shows the $ E_\mathrm{res}$ for all the pixels selected in the on-source. Here the reference waveform comes from the MaxL estimate by the {\tt SEOBNRv4\_ROM} model \cite{Bohe:2016gbl}, implemented in LALSuite \cite{lalsuite} (LALSim version 1.10.0.1), without HMs (we use the same data as in GW190814 paper \cite{Abbott:2020khf}). The time-frequency shape has the chirp-like pattern found in CBCs, but somewhat wider, which corresponds to a deviation of $E_\mathrm{res}$ from the null hypothesis, associated to the $(3,3)$ multipole.

\begin{figure*}[ht!]
\begin{center}
\includegraphics[width=0.5\textwidth]{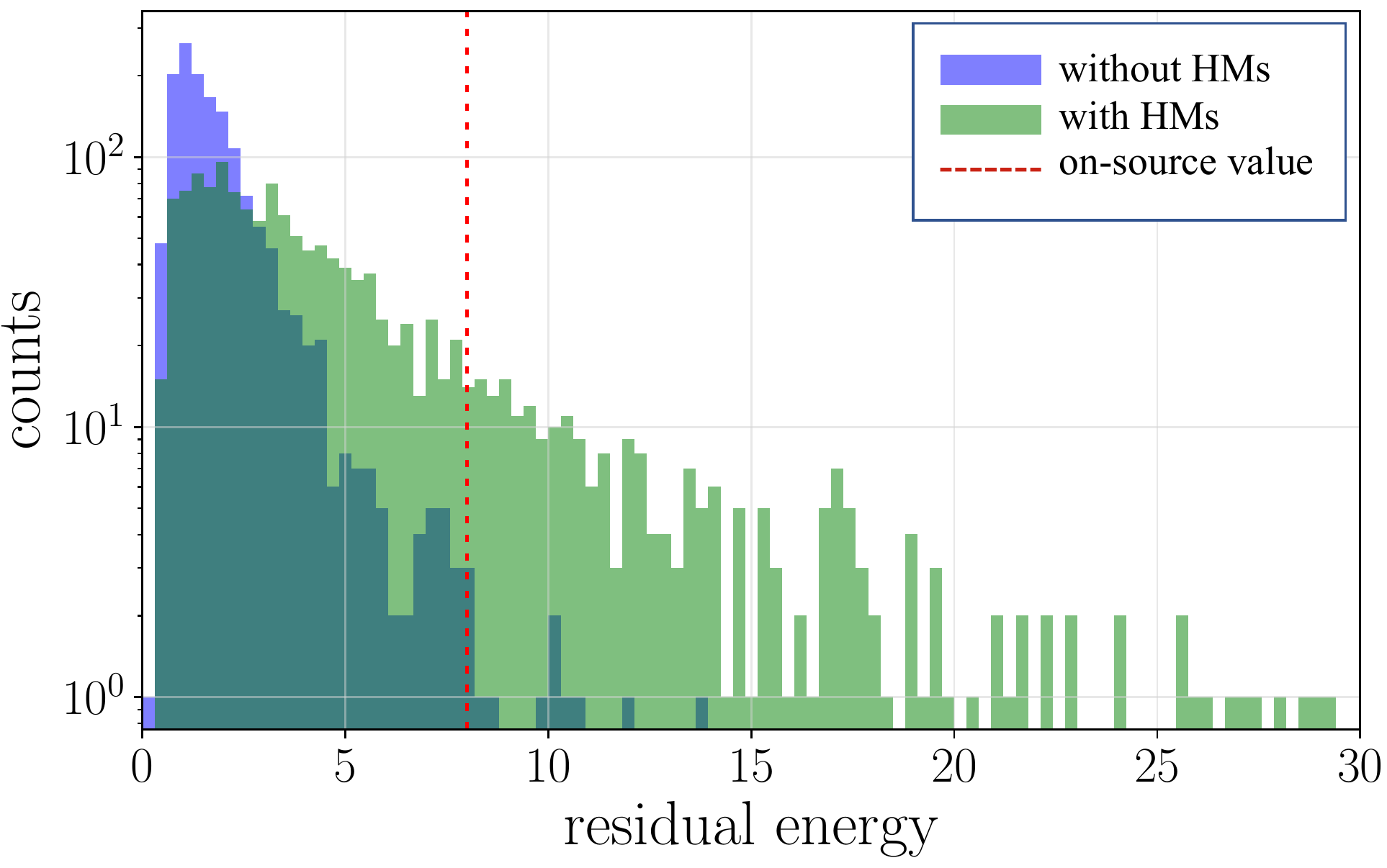}
\end{center}	
\caption{\label{f:distributions} Empirical $E_\mathrm{res}$ distributions of the GW190814 HMs for $\alpha = 1.5$, $\delta \alpha =0.1$, $\Delta t = 0.5$~s, and $\delta t = 0.03$~s. Red vertical line: on-source value. Purple histogram: $E_\mathrm{res}$ distribution for the null hypothesis. Green histogram: $E_\mathrm{res}$ distribution for the model with HMs, from {\tt SEOBNRv4HM\_ROM} injections in off-source. The GW190814 on-source $e_\mathrm{res}$ with respect to the MaxL with HMs is an outlier of the null model, {\tt SEOBNRv4\_ROM} with the \textit{p}-value~=~0.0068, but it is compatible with {\tt SEOBNRv4HM\_ROM} with HMs (\textit{p}-value~=~0.17).}
\end{figure*}

The waveform for the HMs is {\tt SEOBNRv4HM\_ROM} including (2,1), (3,3), (4,4), and (5,5) along with the dominant (2,2) multipole \cite{Cotesta:2018fcv,lalsuite,Cotesta:2020qhw}, in which its HMs can be turned off and simply gives us {\tt SEOBNRv4\_ROM}. Fig.~\ref{f:distributions} gives an example of the empirical distributions of GW190814 for the  slice defined\footnote{The choices of these parameters are discussed in \cite{hom}.} by $\alpha = 1.5$, $\delta \alpha =0.1$, $\Delta t = 0.5$~s, and $\delta t = 0.03$~s. We see that the on-source $E_\mathrm{res}$ with respect to the MaxL with HMs is an outlier of the null hypothesis with the \textit{p}-value~=~0.0068, but it is compatible with {\tt SEOBNRv4HM\_ROM} (the waveform including HMs) with \textit{p}-value~=~0.17.

\begin{figure*}[ht!]
\begin{center}
\includegraphics[clip,width=0.5\textwidth]{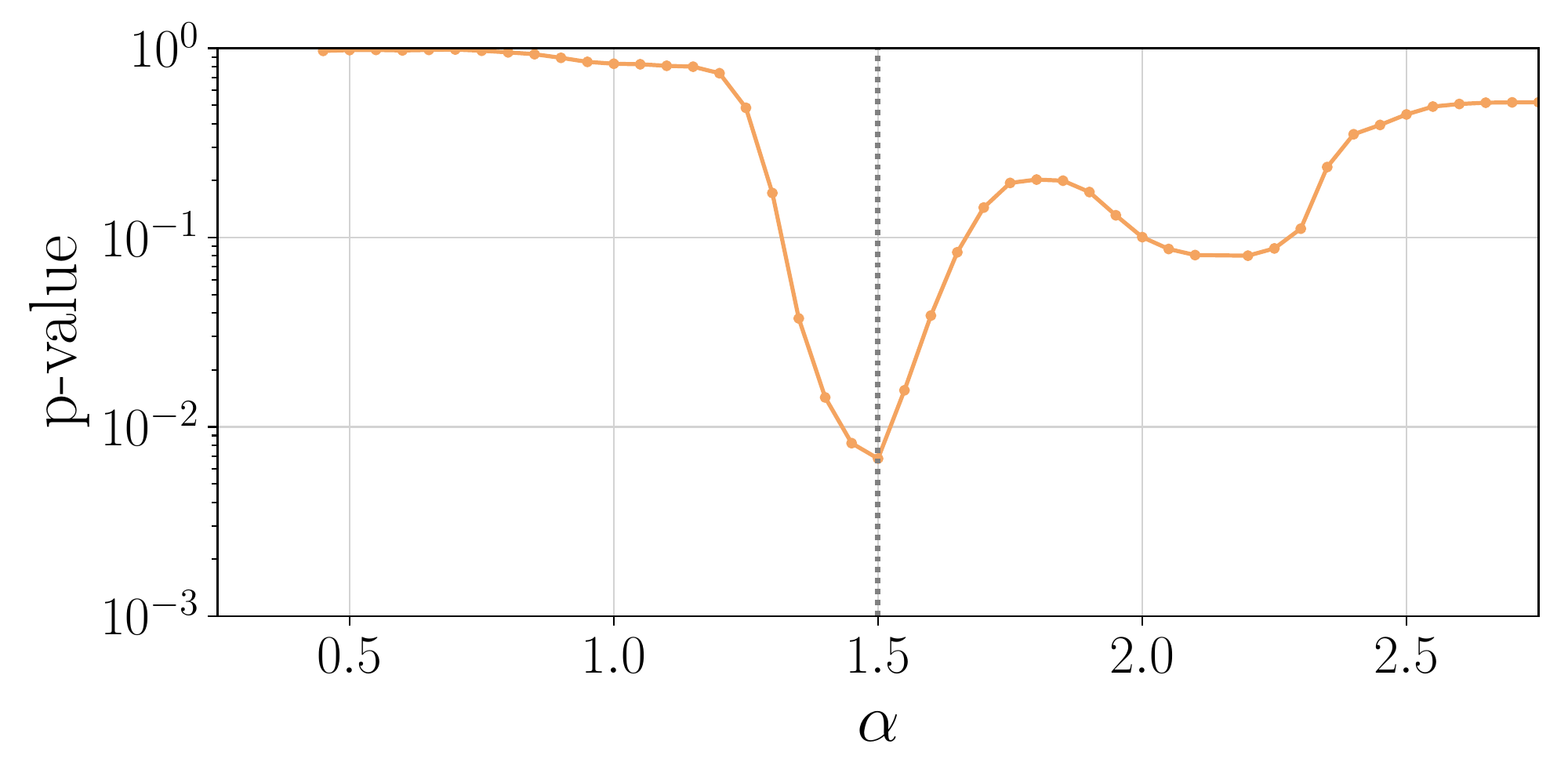}
\end{center}	
	\caption{\label{f:GW190814-pvalue} Plot of \textit{p}-value vs. $\alpha$ for the null hypothesis of GW190814. The \textit{p}-value drops at $\alpha =1.5$ mode, corresponding to the $(3,3)$ multipole. We notice that the quadrupole fluctuation ($\alpha=1$) is consistent with the null hypothesis. {This figure is a deeper version (wider $\alpha$-range and twice of the datapoints) of the lower panel in c.f. Fig. 7 in \cite{Abbott:2020khf}.}}
\end{figure*}

Moreover, we also study other modes. Fig.~\ref{f:GW190814-pvalue} gives us the \textit{p}-value for the null hypothesis for several $\alpha$ values \cite{Abbott:2020khf}. The \textit{p}-value drops at $\alpha =1.5$ (corresponding to $(3,3)$ mode), while the \textit{p}-values are larger for other $\alpha$s that the null hypothesis cannot be rejected. Additionally, cWB does not detect significant $E_\mathrm{res}$ at  $\alpha \sim 1$, implying that the dominant $(2,2)$ multipole (quadrupole) of the {\tt SEOBNRv4HM\_ROM} is consistent with the {\tt SEOBNRv4\_ROM}.

\section{Conclusions}
\label{sec:remarks}

We have given a timely introduction of our method regarding the detection of GW multipoles; the details will be provided in a forthcoming paper \cite{hom}. In particular, we highlight the GW190814 analysis. The paper \cite{hom} is actually the extension of our previous work in \cite{PhysRevD.100.042003}. More details of the Receiver Operating Characteristic curves regarding the choices of $\delta \alpha$ as well as the analysis on GW190412 can be found in that forthcoming paper \cite{hom}.

\ack
cWB makes use of data, software and/or web tools obtained from the Gravitational Wave Open Science Center \cite{gwosc}, a service of LIGO Laboratory, the LIGO Scientific Collaboration and the Virgo Collaboration. LIGO is funded by the U.S. National Science Foundation. Virgo is funded by the French Centre National de Recherche Scientifique (CNRS), the Italian Istituto Nazionale della Fisica Nucleare (INFN) and the Dutch Nikhef, with contributions by Polish and Hungarian institutes. The authors are grateful for computational resources provided by the LIGO Laboratory and supported by National Science Foundation Grants PHY-0757058 and PHY-0823459. 


\end{document}